\documentclass[5p]{elsarticle}

 \makeatletter
 \def\ps@pprintTitle{%
  \let\@oddhead\@empty
  \let\@evenhead\@empty
  \def\@oddfoot{}%
  \let\@evenfoot\@oddfoot}
 \makeatother

\biboptions{numbers,sort&compress}

\usepackage{lineno}
\modulolinenumbers[5]
\usepackage[utf8]{inputenc}
\usepackage[hyphens]{url}
\usepackage{mathtools}
\usepackage{amssymb}
\usepackage{graphicx}
\usepackage[colorlinks=true,linkcolor=blue,urlcolor=blue,citecolor=blue,anchorcolor=blue]{hyperref}
\usepackage{multirow}
\usepackage{color}
\usepackage{dsfont}
\usepackage{algorithm}
\usepackage{algpseudocode}
\usepackage{textcomp}

\usepackage{pdfpages}

\newcommand{\pr}[1]{\left(#1 \right)} 
\newcommand{\cbrace}[1]{\left\{#1 \right\}} 
\renewcommand{\v}[1]{\ensuremath{\mathbf{#1}}} 
\newcommand{\gv}[1]{\ensuremath{\mbox{\boldmath$ #1 $}}} 
\newcommand{\avg}[1]{\left< #1 \right>} 

\let\basetop=\top
\renewcommand{\top}{{}^\basetop \!}
\newcommand{\dx}{\: \mathrm{d}}

\journal{Computer Physics Communications}

\begin{document}

\begin{frontmatter}

\title{Efficient sampling of spreading processes on complex networks using a composition and rejection algorithm}

\author[ulaval,cimmul]{Guillaume St-Onge}
\ead{guillaume.st-onge.4@ulaval.ca}

\author[ulaval,cimmul]{Jean-Gabriel Young}
\author[ulaval,uvm]{Laurent H\'ebert-Dufresne}
\author[ulaval,cimmul]{Louis J. Dub\'e}

\address[ulaval]{D\'epartement de Physique, de G\'enie Physique, et d'Optique, 
Universit\'e Laval, Qu\'ebec (Qu\'ebec), Canada, G1V 0A6}
\address[cimmul]{Centre interdisciplinaire de mod\'elisation math\'ematique de l\textquotesingle Universit\'e Laval, Qu\'ebec (Qu\'ebec), Canada, G1V 0A6}
\address[uvm]{Department of Computer Science and Vermont Complex Systems Center, University of Vermont, Burlington, VT 05401, USA}

\begin{abstract}
Efficient stochastic simulation algorithms are of paramount importance to the study of spreading phenomena on complex networks. Using insights and analytical results from network science, we discuss how the structure of contacts affects the efficiency of current algorithms. We show that algorithms believed to require $\mathcal{O}(\log N)$ or even $\mathcal{O}(1)$ operations per update---where $N$ is the number of nodes---display instead a polynomial scaling for networks that are either dense or sparse and heterogeneous. This significantly affects the required computation time for simulations on large networks. To circumvent the issue, we propose a node-based method combined with a composition and rejection algorithm, a sampling scheme that has an average-case complexity of $\mathcal{O} [\log(\log N)]$ per update for general networks. This systematic approach is first set-up for Markovian dynamics, but can also be adapted to a number of non-Markovian processes and can enhance considerably the study of a wide range of dynamics on networks.
\end{abstract}


\end{frontmatter}


\section{Introduction}

Stochastic processes in a discrete state space are useful models for many natural and human-related complex systems. Considering that complex and heterogeneous connectivity patterns form the backbone of these systems, the study of dynamical processes on networks has grown in popularity in the last decades. Spreading processes are a prime example \cite{Pastor-Satorras2015,Wang2017,Kiss2017}. They are used to model a wide range of phenomena related to propagation among a population. While disease epidemics \cite{Anderson1991} is the most commonly studied process, other important examples include social contagions \cite{Morgan2018,Lehmann2018} and even beneficial epidemics \cite{Berdahl2016}.

Many analytical approaches have been developed to study the outcome of spreading processes on complex networks, using mean field \cite{Boguna2002,Moreno2002,VanMieghem2009,Barrat2008}, moment closure \cite{Eames2002,Mata2014,Sharkey2015,St-Onge2018}, percolation mapping \cite{Newman2002,Kenah2007,Parshani2010} and message passing techniques \cite{Karrer2010,Shrestha2015} to name a few (see Refs.~\cite{Pastor-Satorras2015,Wang2017,Kiss2017} for recent overviews). However, most results hold only for random networks and tree-like structures, or stand as approximations for general networks. This undeniably contributes to our understanding of real systems, but any conclusions drawn from these approaches need to be supported by exact, robust, numerical simulations.

However, an important limitation of numerical simulations is certainly their computational cost, which generally increases with the size of the system. Spreading processes are a class of nonequilibrium models that exhibit a phase transition in the thermodynamic limit (infinite size systems). This particular feature has motivated the need for simulation algorithms that can handle network with very large number of nodes $N$. This is especially true for the study of anomalous phenomena, such as the localization of the epidemic \cite{Goltsev2012,Castellano2012,Pastor-Satorras2018,Liu2019}, Griffith phases and smeared phase transitions \cite{Munoz2010,Odor2014,Mata2015,Cota2016,Cota2018,St-Onge2018}, typically observed in large-size networks. Therefore, the development of stochastic simulation algorithms that can efficiently tackle large networks of varied structures is unescapable, although it offers a considerable, albeit fascinating, computational challenge. It is also critical to the foundations of the field of spreading processes, one of the cornerstones of network science.

One standard formulation of spreading on networks is in terms of a continuous-time stochastic process. A widely used numerical approach involves a temporal discretization using a finite time step. Despite its simplicity, this is both computationally inefficient---requiring $\mathcal{O}(N)$ operations per time step---and prone to discretization errors \cite{Fennell2016}. Instead, the state-of-the-art methods are based on the Doob-Gillespie's algorithm \cite{Gillespie1976}, which produces statistically exact sequences of states. Many studies have been dedicated to the improvement of its efficiency \cite{Gibson2000,Slepoy2008,Goutsias2013,Yates2013}, and from these ideas have emerged some implementations for spreading processes \cite{Vestergaard2015,Kiss2017,Cota2017,Masuda2018,DeArruda2018}. However, these approaches can still be inefficient for certain structures, for instance dense or sparse and heterogeneous networks.

In this work, we propose an efficient method for the simulation of Markovian spreading processes that builds on the Doob-Gillespie's algorithm. It hinges on a composition and rejection scheme \cite{Slepoy2008} that requires only $\mathcal{O}[\log (\log N)]$ operations per update for general networks. We demonstrate that it provides a formidable improvement compared to other algorithms whose computation time can scale polynomially with the network size. An implementation of the algorithm is available online \cite{St-Onge2018repo}.

\section{Markovian spreading processes on networks}

For the sake of simplicity, we consider simple graphs of $N$ nodes and $M$ edges, although the methods discussed can all be applied to directed, weighted, and multi-layer networks. We consider spreading processes defined as continuous Markov processes with discrete states $\gv{x}$. We can focus on canonical epidemiological models to understand the methods---other compartmental models can also be accomodated, for instance by adding accessible states for the nodes, but numerical approaches are similar. Let us therefore denote the state of each node $i$ as $x_i \in \{S, I, R \}$, susceptible, infected or recovered respectively. If infected node $i$ and susceptible node $j$ are connected, node $i$ transmits the disease to node $j$ at rate $\lambda$. Node $i$ spontaneously recovers at rate $\mu$, and becomes recovered if it develops immunity against the spreading disease, or susceptible otherwise. From this framework, we distinguish two models with different properties in the limit $t \to \infty$.

\begin{figure}[tb]
	\centering
	\includegraphics[width = 0.5\textwidth]{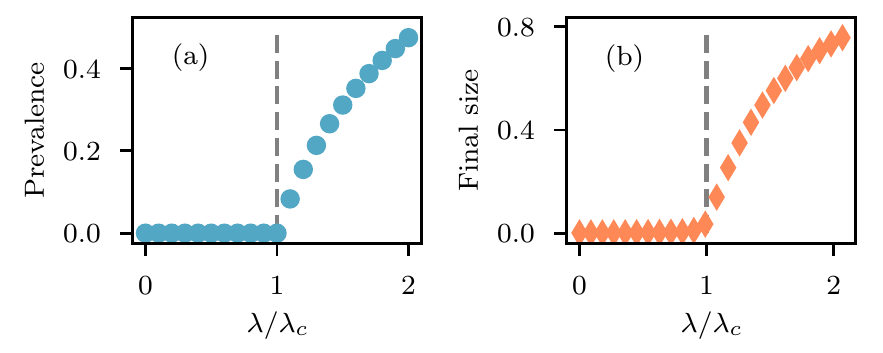}
	\caption{Typical phase transition of the order parameters for the two canonical models of spreading. The recovery rate is fixed to $\mu = 1$ while the transmission rate $\lambda$ is varied. Simulations were performed using the composition and rejection algorithm (see Sec.~\ref{sec:composition_rejection_sampling}) on a single realisation of the $G(N,M)$ random graph ensemble \cite{Erdos1959}, with $N = 10^5$ and $M = 5\times 10^5$. \textbf{(a)} Average prevalence in the endemic state for the SIS model. \textbf{(b)} Average final size for an outbreak of the SIR model, starting with an initial infected node density of $10^{-3}$. In both cases, the error bars associated to the standard deviation are hidden by the markers.}
	\label{fig:phase_diagram}
\end{figure}

\paragraph{SIS model} Infected nodes become directly susceptible after recovery (they do not develop immunity). For finite size networks, the absorbing state with all nodes susceptible, where the system remains trapped, is eventually visited. However, for infinite size networks\footnote{In practice, this is observed for large (finite size) networks.}, there exists some threshold $\lambda_c$ for fixed $\mu$ such that the system reaches an endemic state, where a stationary density of the nodes remains infected on average, for all $\lambda > \lambda_c$ [see Fig.~\ref{fig:phase_diagram}(a)]. This stationary density of infected nodes is hereafter referred as the \emph{prevalence}.

\paragraph{SIR model} Recovered nodes develop immunity. After a certain time, all nodes are either susceptible or recovered. Similarly to the SIS model, there exists a threshold value $\lambda_c$ above which even an infinitesimal inital density of infected nodes will ultimately lead to a macroscopic number of recovered nodes that scales with network size [see Fig.~\ref{fig:phase_diagram}(b)]. The final density of recovered nodes is hereafter referred as the \emph{final size}. \\

Threshold estimates are provided in \ref{app:random_network} for uncorrelated random networks with an arbitrary degree distribution.

\section{Exact simulation algorithms \label{sec:simulation_algorithms}}

Spreading processes can be decomposed into a set $\Omega(\gv{x}) = \{ u_{\alpha} \}$ of independent processes for each state $\gv{x}$ of the system. A process $u_\alpha$ can be the transmission of a disease from node $i$ to $j$ or the recovery of an infected node. We can distinguish the set of transmission processes $\Omega_T(\gv{x})$ and the set of recovery processes $\Omega_R(\gv{x})$ such that $\Omega_T \cap \Omega_R = \emptyset$  and $\Omega_T \cup \Omega_R = \Omega $. To establish a coherent framework for the stochastic simulation algorithms, we consider that the execution of a process is an event that \emph{may} change the state $\gv{x}$ of the system, but it is not required to do so. For instance, the transmission of a disease from node $i$ to $j$ is a process that implies $x_j \mapsto I$ if $x_j = S$. However, if $x_j \in \{I,R\}$ prior to the transmission, then the state of the system stays the same. Considering or not the transmission of a disease between two infected nodes as an ongoing process is only an algorithmic decision; from our definition, an infected node transmits the disease to all of its neighbors irrespectively of their states.

Since every possible transmission and recovery for a state $\gv{x}$ is an independent process $u_\alpha$ with rate $w_\alpha \in \cbrace{\lambda, \mu}$, the total rate of events is defined as
\begin{align}\label{eq:total_rate}
	W(\gv{x}) = \sum_{u_\alpha \in \Omega(\gv{x})} w_\alpha \;.
\end{align}
The inter-event time $\tau(\gv{x})$ before the execution of any process is exponentially distributed
\begin{align}\label{eq:global_inter-event_time}
	P(\tau| \gv{x}) =  \exp(-W(\gv{x}) \tau) \mathrm{d} t,
\end{align}
and the executed process is chosen proportionally to its \emph{propensity}, i.e. its rate $w_\alpha$
\begin{align}\label{eq:process_distribution}
	P(u_\alpha|\v{x}) = \frac{w_\alpha}{W(\gv{x})} \;.
\end{align}

In this section, we introduce and discuss different numerical methods to sample sequences of states based on the previous equations. To compare the methods and give insights on their behavior for large system, we will provide expressions for the theoretical number of operations required on average to perform a state transition, i.e. $\gv{x} \mapsto \gv{x}'$ with $\gv{x}' \neq \gv{x}$. The expressions are \emph{upper bounds} to the number of operations required on average and are expressed using the big $\mathcal{O}$ notation.

\subsection{Doob-Gillespie's algorithm (direct method)}

The direct method consists of the determination of the next process $u_\alpha$ and the time at which it will be executed. It can be summarized by these steps :
\begin{enumerate}
	\item Determine the inter-event time $\tau(\gv{x})$ using Eq.~\eqref{eq:global_inter-event_time}.
	\item Determine the executed process $u_\alpha$ using Eq.~\eqref{eq:process_distribution}.
	\item Update the state $\gv{x} \mapsto \gv{x}'$ according to the chosen process $u_\alpha$ (if $\gv{x}' \neq \gv{x}$), and update the time $t \mapsto t + \tau(\gv{x})$.
	\item Update, create, or remove processes resulting from the execution of process $u_\alpha$. 
	\item Update $W(\gv{x})$ and $P(u_\alpha|\v{x})$, then return to step 1.
\end{enumerate}
There have been various implementations of this direct procedure that have been discussed for spreading processes on static networks \cite{Fennell2016,Cota2017,Kiss2017,Masuda2018}. 

For general networks, the number of ongoing processes is $\mathcal{O}(M + N)$. Efficient implementations typically use a binary-tree data structure to store the processes and keep updated $P(u_\alpha |\gv{x})$ \cite{Gibson2000,Masuda2018}. This allows the insertion and the retrieval of processes in $\mathcal{O}(\log N)$ operations. The costly processes involve the infection of a degree $k_I$ node. To perform this update, the required number of operations is
\begin{align}
	\mathcal{O}\pr{k_I \log N} \;
\end{align}
and is associated with the storage of future transmission processes. 

A fact often overlooked is that the degree $k_I$ of a newly infected node can be large on average. For dense networks, it scales as $k_I \sim N$ ; for sparse and heterogeneous networks near the phase transition, $k_I \sim N^\xi$ on average with $0 \leq \xi \leq 1$ (see \ref{app:random_network}). This is problematic, since phase transitions are central to many studies.

\subsection{Node-based method \label{sec:node-based}}

Another type of implementation has been suggested \cite{Cota2017,DeArruda2018}. To sample among the infection processes, one selects a node $i$ among the infected nodes, proportionally to its degree $k_i$, then selects one of its neighbors $j \in \partial i$ randomly and infects it if $x_j = S$. The sampling of recovery processes is simply done by selecting an infected node uniformly. Using this scheme, one does not have to look through all the neighbors of an infected node to check for new processes $u_\alpha$, associated with the $\mathcal{O}(k_I)$ operations in the direct method.

We introduce what we call the \emph{node-based} method in the spirit of this scheme. The idea is to regroup the propensity of all processes associated with a node into a single propensity in order to prevent their enumeration. The probability of selecting a process $u_\alpha$ performed by node $i$ is then factorized as $P(u_\alpha,i| \gv{x}) = P(u_\alpha| i, \gv{x})P(i| \gv{x}) $. The probability of selecting a node is written
\begin{align}\label{eq:per_node_propensity_distribution}
	P(i|\gv{x}) = \frac{\omega_i}{W(\gv{x})}\;,
\end{align}
where 
\begin{align} \label{eq:individual_propensity}
\omega_i(\gv{x}) &= \delta(x_i,I) (\lambda k_i + \mu) \;
\end{align}
is the total propensity for node $i$ to be involved in a transmission or recovery event, $\delta(a,b)$ is the Kronecker delta and $k_i$ is the degree of node $i$. The probability of selecting a process $u_\alpha$ performed by node $i$ is formally written
\begin{align}
P(u_\alpha|i, \gv{x}) = \frac{w_\alpha}{\omega_i(\gv{x})} \;.
\end{align}
In practice, we use the probability that the chosen process is a transmission
\begin{align}
	P(u_\alpha \in \Omega_T| i, \gv{x}) = \frac{\lambda k_i}{ \lambda k_i + \mu} \;,
\end{align}
to select the type of process. If it is a transmission, we select a random neighbor and infect it; if it is a recovery, the node becomes susceptible or recovered depending on the spreading model used. Note that the total rate of events [Eq.~\eqref{eq:total_rate}] has an equivalent definition in terms of the propensities of the nodes, namely
\begin{align}\label{eq:total_rate_nb}
	W(\gv{x}) = \sum_{i=1}^n \omega_i(\gv{x})\;.
\end{align}

An important property of this scheme in the context of spreading processes is that Eq.~\eqref{eq:individual_propensity} is only dependent on the state of the node $i$. Hence, when the state of a node $i$ changes due to the execution of a process, we do not need to update the propensity $\omega_j$ of neighboring nodes $j\in \partial i$, but only $\omega_i$. We discuss two implementations for this method.

\subsubsection{Rejection sampling \label{sec:rejection_sampling}}

\begin{figure}[tb]
	\centering
	\includegraphics[width = 0.5\textwidth]{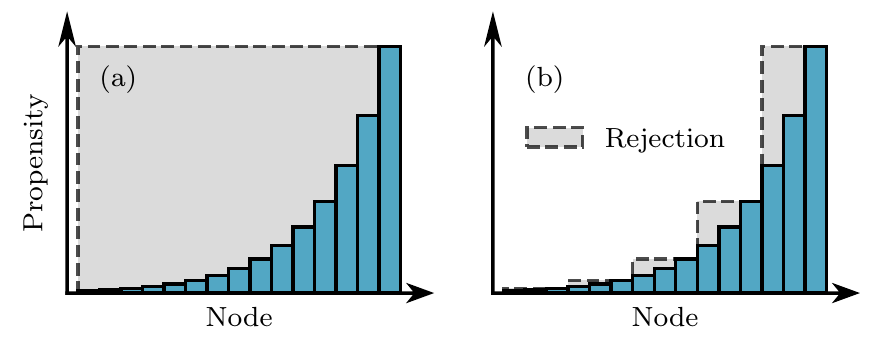}
	\caption{ Rejection area (gray) compared to the acceptance area (blue), for the selection of a node with the node-based method. \textbf{(a)} Rejection sampling. \textbf{(b)} Composition and rejection sampling.}
	\label{fig:rejection_area}
\end{figure}

A simple approach to select a node $i$ according to $P(i|\gv{x})$ is to use rejection sampling. First, one needs to determine the maximal propensity possible $\omega_\mathrm{max}$ for a node, in this case being
\begin{align}\label{eq:max_propensity}
	\omega_\mathrm{max} = \lambda k_\mathrm{max}+\mu\;,
\end{align}
where $k_\mathrm{max}$ is the maximal degree of a node in the network. Second, one needs to keep updating an array of pairs of elements $\mathcal{A}$, the pairs being node labels and their propensity $(i,\omega_i)$---we keep only pairs with non-zero propensities $\omega_i$. Finally, one draws a pair $(i,\omega_i)$ at random from $\mathcal{A}$ and accepts it with probability
\begin{align}\label{eq:acceptance_probability}
	P[(i,\omega_i) | \omega_\mathrm{max}] = \frac{\omega_i}{\omega_\mathrm{max}} \;.
\end{align}
The complete procedure for the execution of a process using the node-based method with rejection sampling is presented in Algorithm \ref{alg:node-based_rejection}. Although the SIS model was considered, one only needs to change step 19 by $S \mapsto R$ for it to be compliant with the SIR model.

\begin{algorithm}[tb]
\caption{Node-based method with rejection sampling for an update of the SIS model}
\label{alg:node-based_rejection}
  \textbf{Input:} $\gv{x}$, $\mathcal{A}$, $W$ \\
  \textbf{Output:} $\tau$ and updated input
\begin{algorithmic}[1]
	\State choose $r_1 \in [0,1)$ uniformly at random
	\vspace{0.1cm}
	\State $\tau \gets -\frac{\ln r_1}{W}$
	\vspace{0.1cm}
	\State select $(i,\omega_i) \in \mathcal{A}$ uniformly at random
	\State choose $r_2 \in [0,1)$ uniformly at random
	\vspace{0.1cm}
	\If{$r_2 < \frac{\omega_i}{\omega_\mathrm{max}}$}
	\vspace{0.1cm}
		\State choose node $i$
	\Else
		\State go back to step 3
	\EndIf
	\State choose $r_3 \in [0,1)$ uniformly at random
	\vspace{0.1cm}
	\If{$r_3 < \frac{\lambda k_i}{\lambda k_i + \mu}$} \Comment{transmission process}
	\vspace{0.1cm}
		\State choose $j \in \partial i$ uniformly at random
		\If{$x_j = S$}
			\State $x_j \gets I$
			\State $\mathcal{A} \gets \mathcal{A}\cup \{(j,\omega_j) \}$
			\State $W \gets W + \omega_j$
		\EndIf
	\Else \Comment{recovery process}
		\State $x_i \gets S$
		\State $\mathcal{A} \gets \mathcal{A} \setminus \{(i,\omega_i) \}$
		\State $W \gets W - \omega_i$
	\EndIf
\end{algorithmic}
\end{algorithm}

It is straightforward to verify that most steps are $\mathcal{O}(1)$ in Algorithm \ref{alg:node-based_rejection}, except possibly the rejection sampling portion (steps 3 to 9). The average acceptance probability for a state $\gv{x}$ is $\avg{\omega_i|I;\gv{x}}/\omega_\mathrm{max}$, where
\begin{align}
\avg{\omega_i|I;\gv{x}} = \lambda \avg{k|I;\gv{x}}  + \mu \;,
\end{align}
is the average propensity for infected nodes in state $\gv{x}$ and $\avg{k|I;\gv{x}}$ stands for the average degree of infected nodes in state $\gv{x}$. Since the number of required operations follows a geometric distribution, the average number of operations for an update from a state $\gv{x}$ is 
\begin{align}\label{eq:complexity_rejection}
	\mathcal{O}(\omega_\mathrm{max}/\avg{\omega_i|I;\gv{x}}) \;.
\end{align}

For networks with a homogeneous degree distribution, we have $\avg{\omega_i|I;\gv{x}} \sim \omega_\mathrm{max}$ for all states, leading to a small number of operations. However, this rejection sampling scheme is vulnerable to cases where the propensity ranges over multiple scales---more specifically, when $\avg{\omega_i|I;\gv{x}} \ll \omega_\mathrm{max}$ for typical states $\gv{x}$. This happens when $k_\mathrm{max}$ is large, which is common for heterogeneous degree distribution, such as $P(k) \sim k^{-\gamma}$ (see \ref{app:random_network}). To circumvent this, we must update and sample efficiently from a heterogeneous distribution of propensities $P(i|\gv{x})$ [Eq.~\eqref{eq:per_node_propensity_distribution}].

\subsubsection{Composition and rejection for multiscale propensity \label{sec:composition_rejection_sampling}}

The high rejection probability of the rejection sampling for heterogeneous propensities is illustrated by the gray portion in Fig.~\ref{fig:rejection_area}(a). The problem is the uniform proposal distribution on $\mathcal{A}$ (step 3 of Algorithm \ref{alg:node-based_rejection}), which is a poor choice for a heterogenous distribution $P(i|\gv{x})$. To improve upon rejection sampling, we need an algorithm that \emph{systematically} constructs a proposal distribution that upper bounds the rejection probability, as illustrated in Fig.~\ref{fig:rejection_area}(b).

We propose to use a method of \emph{composition and rejection}, similarly to Ref.~\cite{Slepoy2008} for biochemical reaction networks, inspired by Ref.~\cite{Devroye1986}. It is also a direct improvement over the 2-group method proposed in Ref.~\cite{Cota2017}. The idea is to create a partition of nodes $\mathcal{P} =\{ g_\beta \}$ with similar propensities $\omega > 0$, with $\beta \in \{1, 2, \dots, q \}$. Once a node $i$ gets infected (or more generally gets a propensity $\omega_i > 0$), it is assigned to a group $g_\beta$---in our implementation, the pair $(i, \omega_i)$ is stored in an array $\mathcal{A}_\beta$. The probability to select a node $i$ is then factorized as $P(i| \gv{x}) = P(i| g_\beta,\gv{x}) P(g_\beta|\gv{x})$. 

\begin{figure}[tb]
	\centering
	\includegraphics[width = 0.49 \textwidth]{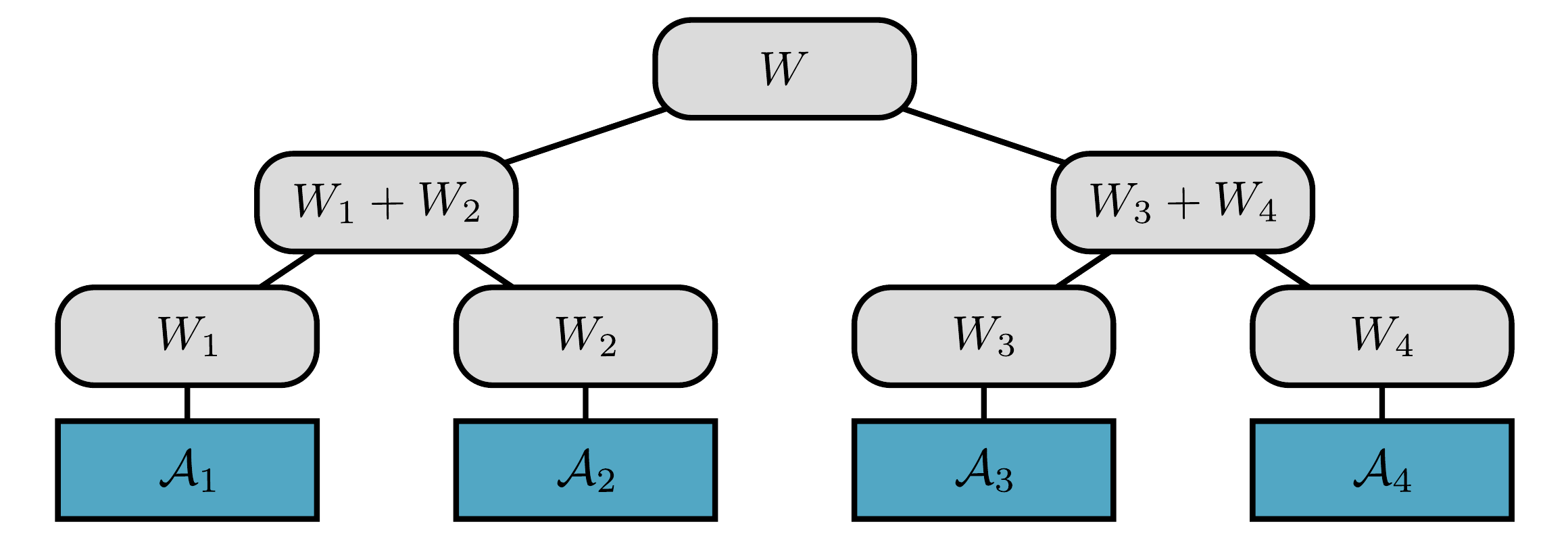}
	\caption{Decision tree $\mathcal{T}$ used to select the arrays $\cbrace{\mathcal{A}_\beta}$.}
	\label{fig:binary_tree}
\end{figure}

The probability of selecting a group $g_\beta$ is
\begin{align}
	P(g_\beta| \gv{x}) = \frac{W_\beta(\gv{x})}{W(\gv{x})} \;,
\end{align}
where ${W_\beta(\gv{x}) \equiv \sum_{i \in g_\beta} \omega_i(\gv{x})}$ is the propensity associated to the group of nodes. In practice, to select an array $\mathcal{A}_\beta$ proportionally to $W_\beta$, we implement a binary decision tree $\mathcal{T}$ as illustrated in Fig.~\ref{fig:binary_tree}, where each leaf points to an array $\mathcal{A}_\beta$. Starting from the root, it takes $\mathcal{O}(\log q)$ operations to choose one of the leaves. Once a process $u_\alpha$ is chosen and executed, we update the array $\mathcal{A}_\beta$, the propensity $W_\beta$ and recursively the parent values in the tree---one notes that the root value is in fact $W(\gv{x})$. Again, $\mathcal{O}(\log q)$ operations are needed for this task. 

The probability of selecting a node within a group is
\begin{align}
	P(i|g_\beta,\gv{x}) = \frac{\omega_i(\gv{x})}{W_\beta(\gv{x})} \;.
\end{align}
If the partition $\mathcal{P}$ is wisely chosen, nodes are selected efficiently using rejection sampling, replacing $\omega_\mathrm{max}$ in Eq.~\eqref{eq:acceptance_probability} by a group specific maximal propensity $\omega_{\beta\mathrm{max}}$.

A systematic approach to construct the partition $\mathcal{P}$ is to impose a minimum acceptance probability, say one half. This leads to an average number of operations upper-bounded by 2. Let us define the minimal propensity as
\begin{align}
	\omega_\mathrm{min} = \lambda k_\mathrm{min} + \mu \;,
\end{align}
where $k_\mathrm{min}$ is the minimal degree in the network. We impose that the $\beta$-th group allows nodes with propensity $\omega \in [2^{\beta-1}\omega_\mathrm{min}, 2^\beta\omega_\mathrm{min})$, except for the last group allowing $\omega \in [2^{q-1}\omega_\mathrm{min}, \omega_\mathrm{max}]$. The group specific maximal propensity is thus
\begin{align}
	\omega_{\beta\mathrm{max}} = \mathrm{min}\pr{2^\beta \omega_\mathrm{min}, \omega_\mathrm{max}} \;,
\end{align}
and the number of group required is
\begin{align}\label{eq:number_of_groups}
 	q = \mathrm{max} \left(1, \left\lceil \log_2 \left( \frac{\omega_\mathrm{max}}{ \omega_\mathrm{min} }\right) \right\rceil \right)\;.
 \end{align}
The complete procedure for the execution of a process using the node-based method with composition and rejection sampling is presented in Algorithm \ref{alg:node-based_composition-rejection}.

\begin{algorithm}[tb]
\caption{Node-based method with composition and rejection sampling for an update of the SIS model}
\label{alg:node-based_composition-rejection}
  \textbf{Input:} $\gv{x}$, $\mathcal{T}$ \\
  \textbf{Output:} $\tau$ and updated input
\begin{algorithmic}[1]
	\State get $W$ from the root of $\mathcal{T}$
	\State choose $r_1 \in [0,1)$ uniformly at random
	\vspace{0.1cm}
	\State $\tau \gets -\frac{\ln r_1}{W}$
	\vspace{0.1cm}
	\State choose $r_2 \in [0,1)$ uniformly at random
	\State select $\mathcal{A}_\beta$ from $\mathcal{T}$ using $r_2$
	\State select $(i,\omega_i) \in \mathcal{A}_\beta$ uniformly at random
	\State choose $r_3 \in [0,1)$ uniformly at random
	\vspace{0.1cm}
	\If{$r_3 < \frac{\omega_i}{\omega_{\beta\mathrm{max}}}$}
	\vspace{0.1cm}
		\State choose node $i$
	\Else
		\State go back to step 6
	\EndIf
	\State choose $r_4 \in [0,1)$ uniformly at random
	\vspace{0.1cm}
	\If{$r_4 < \frac{\lambda k_i}{\lambda k_i + \mu}$} \Comment{transmission process}
	\vspace{0.1cm}
		\State choose $j \in \partial i$ uniformly at random
		\If{$x_j = S$}
			\State $x_j \gets I$
			\State $\mathcal{A}_\mathcal{\beta'} \gets \mathcal{A}_\mathcal{\beta'}\cup \{(j,\omega_j) \}$ \Comment{$j \in g_{\beta'}$}
			\State update $\mathcal{T}$
		\EndIf
	\Else \Comment{recovery process}
		\State $x_i \gets S$
		\State $\mathcal{A}_\mathcal{\beta} \gets \mathcal{A}_\mathcal{\beta} \setminus \{(i,\omega_i) \}$
		\State update $\mathcal{T}$
	\EndIf
\end{algorithmic}
\end{algorithm}

For general networks with $\lambda/\mu < \infty$, the ratio of extreme propensities $\omega_\mathrm{max}/\omega_\mathrm{min} = \mathcal{O}(N)$. Therefore, the average-case complexity per update is
\begin{align}
	\mathcal{O}[\log(\log N)] \;.
\end{align}
For networks with a maximal degree $k_\mathrm{max}$ independent of $N$, the complexity is $\mathcal{O}(1)$.

It is worth mentioning that our choice to impose a minimum acceptance probability of one half is not unique : one could impose an acceptance probability of $a^{-1}$ with $a > 1$ and obtain the same average-case complexity. For $a \to 1$, it increases the acceptance probability, but it also increases the number of groups required $q$. Therefore, there is a trade-off to consider when one tries to minimize the required number of operations. We made several trials with $a \neq 2$, but it has never resulted in noticeable improvements for the computation time.

\subsection{Event-driven method}

Another type of approach for the simulation of spreading processes has been considered lately \cite{Kiss2017,DeArruda2018}. The philosophy is similar to the next reaction method \cite{Gibson2000}, originally proposed for the simulation of chemical reaction networks to improve upon the original Doob-Gillespie's algorithm. The principal concept of this scheme is to draw an inter-event time $\tau_\alpha$ for each of the specific process $u_\alpha$, and execute the latter at time $t_\alpha + \tau_\alpha$, where $t_\alpha$ is the absolute time when the process $u_\alpha$ was created. Therefore, one focus on the execution time of each process independently, instead of inferring the global inter-event time $\tau(\gv{x})$ and the first process to be executed among $\Omega(\gv{x})$, as in standard Gillespie method. 

For Markovian dynamics, the inter-event time before the execution of a process $u_\alpha$ is exponentially distributed 
\begin{align}
	P(\tau_\alpha) =  \exp(-w_\alpha \tau_\alpha) \mathrm{d} t_\alpha \;.
\end{align}
However, it is important to stress that this approach can also be applied to non-Markovian spreading processes \cite{Kiss2017,DeArruda2018}.

To store and retrieve the processes efficiently, one can use a priority queue, where the highest priority element corresponds to the process $u_\alpha$ with the lowest absolute execution time $t_\alpha + \tau_\alpha$. Recovery processes are eventually executed and can be stored directly in the priority queue. Transmission processes are stored if the inter-event time for the transmission is smaller than the inter-event time for the recovery of the infected node. Depending on the implementation, one can also verify \emph{a priori} if a neighbor node will already be infected or recovered, and prevent the storage of the transmission process.

To compare this approach with our node-based schemes, we used the implementation provided by Ref.~\cite{Kiss2017,Millerrepo}, called the \emph{event-driven method}, for which a detailed pseudocode is available for both the SIS and the SIR models. The interface is modular and general enough to be used for any networks, a requirement for our benchmark calculations. Let us mention that there are alternative implementations available, for instance the efficient implementation of Ref.~\cite{Holmesrepo} for the SIR model (see Supplementary Material \cite{SM} for a comparison with our algorithm).

As for the standard Gillespie algorithm, a costly process for the event-driven method involves the infection of a degree $k_I$ node, for which 
\begin{align}\label{eq:complexity_ED}
	\mathcal{O}\pr{k_I \log N} \;,
\end{align}
operations are required. While this quantity is probably not the most representative for the average computation time associated with the algorithm [see Fig.~\ref{fig:computation_time_comparison}], we can clearly identify the different impact of the term $k_I$ for dense or sparse and heterogeneous networks.

\section{Efficiency of the stochastic simulation algorithms}

\begin{figure*}[tb]
	\centering
	\includegraphics[width = 0.98\textwidth]{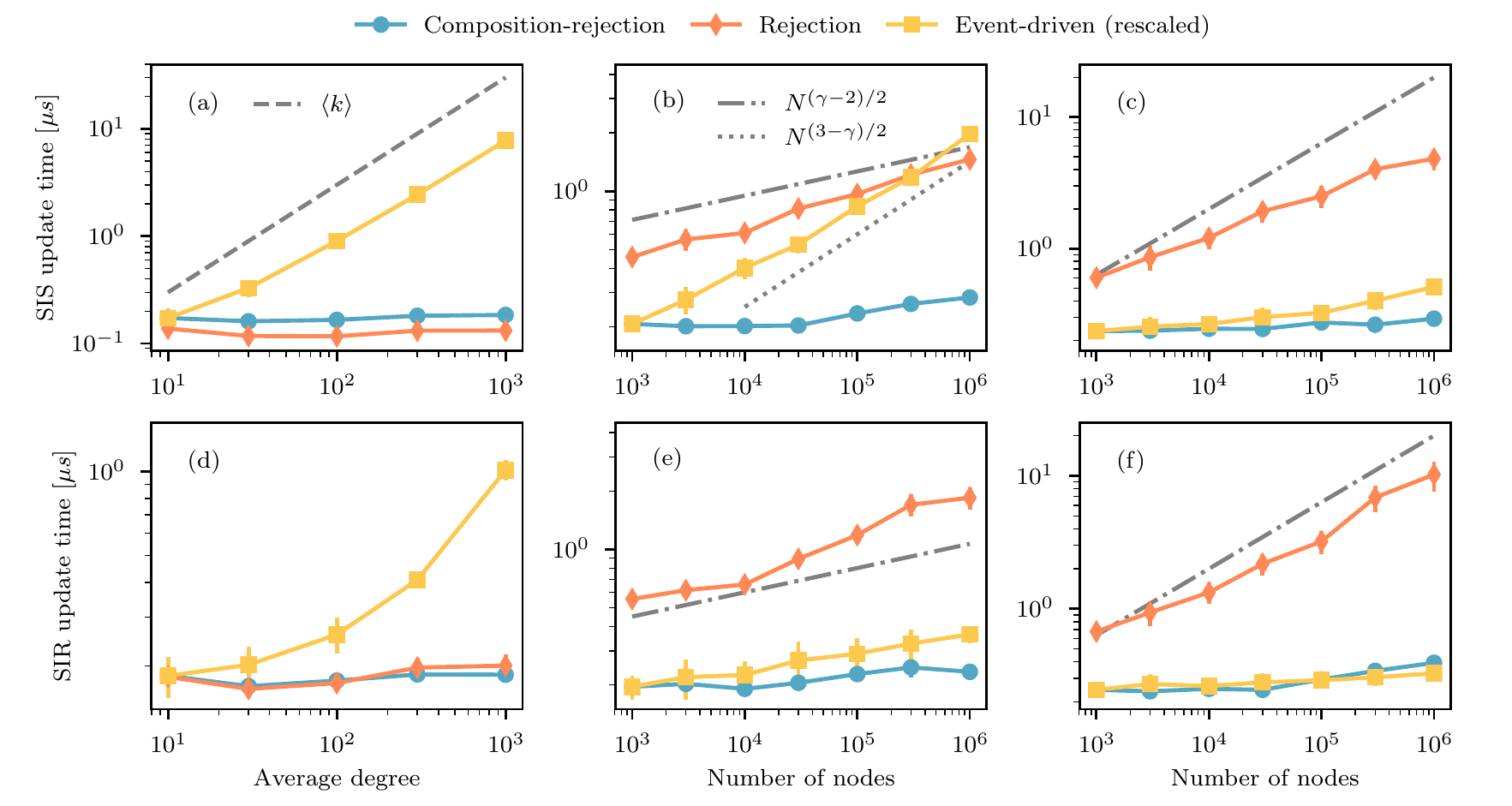}
	\caption{Average computation time for a single update of the state for spreading processes using different algorithms.  Each marker is averaged over 10 simulations on 10 realizations of a random graph ensemble. The computation time of the event-driven method was rescaled to match the first marker of the composition and rejection method. \textbf{(Upper row)} Average over sequences of $10^6$ state transitions for the SIS model in the stationary state. The systems have been thermalized using $10^6$ transitions beforehand. \textbf{(Lower row)} Average over $10^6/N$ complete sequences of the SIR model, starting with an initial infected node density of $10^{-2}$. \textbf{(a)} and  \textbf{(d)} $G(N,M)$ random graphs with fixed number of nodes $N = 10^4$ and different average degree $\avg{k}$. The recovery and transmission rates used are $\mu = 1$ and $\lambda = 1.05 \lambda_c$ . \textbf{(b)} and  \textbf{(e)} Random graphs with an expected degree sequence \cite{Chung2002,Miller2011}. We used a power law expected degree distribution $P(\kappa) \sim \kappa^{-\gamma}$ with $\gamma = 2.25$, $\kappa_\mathrm{min} = 3$ and $\kappa_\mathrm{max} < \avg{k}N^{1/2}$. The recovery and transmission rates used are $\mu = 1$ and $\lambda = 3 \lambda_c$. \textbf{(c)} and  \textbf{(f)} Same as (b) and (e), but with $\gamma = 3$.}
	\label{fig:computation_time_comparison}
\end{figure*}

We have described different schemes for the simulation of spreading processes, with different associated complexity for the update of the state $\gv{x} \mapsto \gv{x}'$. In this section, we show and discuss the impact on the computation time using synthetic networks. In Fig.~\ref{fig:computation_time_comparison} we compile the results of our benchmark calculations, using the SIS model in the stationary state and complete sequences of the SIR model, starting with a small infected node density. We fix the recovery rate $\mu = 1$ without loss of generality. Furthermore, we scale the transmission rate parameter $\lambda = \eta \lambda_c$ according to the underlying threshold $\lambda_c$ of the dynamics. The intent is to always have a similar prevalence (SIS) or final size (SIR) in our simulations as we tune the network structure : as we discuss in \ref{sec:phantom_event_node-based}, different values for the order parameter can affect the expected number of required operations.

Implementations for the two node-based methods---hereafter referred as to \emph{rejection} and \emph{composition and rejection}---are in C++, while the implementation of the event-driven method is in Python \cite{Kiss2017,Millerrepo}. This discrepancy of programming languages causes a multiplicative factor overhead of roughly 100 for the computation times of the event-driven method. To provide a fairer comparison, and since we are mostly interested in the scaling of the algorithms with the size of the network, we have rescaled the computation times obtained with the event-driven method to match the first marker of the composition and rejection method in each panel of Fig.~\ref{fig:computation_time_comparison}. For instance, if $T_\mathrm{ED}(N)$ [$T_\mathrm{CR}(N)$] represents the computation time of the event-driven (composition and rejection) method for different sizes $N$, we perform the transformation $T_\mathrm{ED}(N) \mapsto T_\mathrm{CR}(N_\mathrm{min}) T_\mathrm{ED}(N)/T_\mathrm{ED}(N_\mathrm{min})$, with $N_\mathrm{min} = 10^3$ in Fig.~\ref{fig:computation_time_comparison}. A comparison without rescaling is also available in the Supplementary Material \cite{SM}.

\subsection{Computation time for homogeneous networks}

As a model of homogeneous networks, we used the $G(N,M)$ random graphs ensemble \cite{Erdos1959}. In the limit $N \to \infty$, the degree distribution is a binomial with $\avg{k} = 2M/N$ and the degree of nodes are well represented by the mean value $\avg{k}$. 

We observe in Fig.~\ref{fig:computation_time_comparison}(a) that the average computation time for the event-driven method roughly scales linearly with the average degree, in agreement with Eq.~\eqref{eq:complexity_ED}. For the SIR model in Fig.~\ref{fig:computation_time_comparison}(d), the dependence is less important, except for very large average degree. 

As a comparison, both node-based methods are completely independent of the average degree [Fig.~\ref{fig:computation_time_comparison}(a) and \ref{fig:computation_time_comparison}(d)]. This is in line with the two steps selection procedure for infection processes. A node-based scheme should therefore be privileged whenever one wants to sample networks with large average degree, such as dense networks where $\avg{k} = \mathcal{O}(N)$. 

One also observes that the rejection method is slightly more efficient [Fig.~\ref{fig:computation_time_comparison}(a)] than the composition and rejection method : this is due to the simpler implementation of the rejection sampling, without the need for a composition step. As discussed in Sec.~\ref{sec:composition_rejection_sampling}, the average number of operations $\mathcal{O}(\omega_\mathrm{max}/\avg{\omega_i|I;\gv{x}})$ is only problematic when propensities span multiple scales ; for homogeneous networks, the ratio $\omega_\mathrm{max}/\avg{\omega_i|I;\gv{x}}$ is expected to be $\mathcal{O}(1)$ for all states. 

\subsection{Computation time for heterogeneous networks}

As a model of heterogeneous networks, we used random graphs with an expected degree sequence $\{\kappa_i \}$ \cite{Chung2002,Miller2011}, also called Chung-Lu graphs, where $\kappa_i$ is the expected degree of node $i$. We used sequences drawn from a power-law distribution $P(\kappa) \sim \kappa^{-\gamma}$ to generate heterogeneous networks, with $\gamma = 2.25$ in Fig.~\ref{fig:computation_time_comparison}(b) and Fig.~\ref{fig:computation_time_comparison}(e), and $\gamma = 3$ in Fig.~\ref{fig:computation_time_comparison}(c) and Fig.~\ref{fig:computation_time_comparison}(f).

We observe in Fig.~\ref{fig:computation_time_comparison}(b) that the computation time for the event-driven method scales polynomially with the number of nodes. In Fig.~\ref{fig:computation_time_comparison}(c), the computation time slightly increases, but with a much smaller exponent. This is explained by Eq.~\eqref{eq:complexity_ED}, with 
\begin{align}
	k_I \sim N^{(3-\gamma)/2} \;
\end{align}
near the phase transition (see \ref{app:random_network}). For the SIR model in Fig.~\ref{fig:computation_time_comparison}(e) and \ref{fig:computation_time_comparison}(f), the computation times are less influenced by the size of the network. In this case, the number of required operations predicted in Eq.~\eqref{eq:complexity_ED} by the most costly processes overestimates the average computation time (as also noted for homogeneous networks). 

We observe that the rejection method scales polynomially with the number of nodes as well, but this time the scaling exponent is larger for moderately heterogeneous networks [Fig.~\ref{fig:computation_time_comparison}(c) and \ref{fig:computation_time_comparison}(f)] than for very heterogeneous networks [Fig.~\ref{fig:computation_time_comparison}(b) and \ref{fig:computation_time_comparison}(e)]\footnote{A larger dispersion for the degree distribution implies a more heterogeneous network.}. This is roughly explained by Eq.~\eqref{eq:complexity_rejection}, with $\avg{\omega_i|I;\gv{x}} \sim \avg{k^2}$ (see \ref{app:random_network}) and $\omega_\mathrm{max} \sim k_\mathrm{max}$, leading to
\begin{align}
	\frac{\omega_\mathrm{max}}{\avg{\omega_i|I;\gv{x}}} \sim N^{(\gamma - 2)/2} \;.
\end{align}

Finally, we see that the computation time for the composition and rejection method is, for all practical purposes, independent of the number of nodes.

\section{Concluding remarks}

We have introduced a stochastic simulation algorithm for the simulation of spreading processes on networks, combining a node-based perspective with the efficiency of composition and rejection sampling \cite{St-Onge2018repo}. This algorithm requires $\mathcal{O}[\log (\log N)]$ operations per update, providing a superior scaling (or at worst, equivalent) to the other state-of-the-art methods. It is particularly well suited for the sampling of large and heterogeneous networks, since its average computation time is, for all practical purposes, independent of the network size or the density of edges.

For the SIR model specifically, our benchmark calculations reveal that the event-driven method scales efficiently as well, except for very dense networks. If computation time is a crucial issue, we suggest to compare the speed of our implementation \cite{St-Onge2018repo} with highly optimized implementations of the event-driven method, such as the one provided by Ref.~\cite{Holmesrepo}, or the recent hybrid approach proposed by Ref.~\cite{Grossmann2018}, and choose the one that performs better for the particular case study.

Note that there is a complete branch of literature whose concern is the efficient sampling of the quasistationary distribution of states for processes with an absorbing state, such as the SIS model (see Refs.~\cite{Oliveira2005,Sander2016,Cota2017,Macedo-Filho2018} for instance). Indeed, finite size analysis of the critical phenomenon requires the sampling of sequences that do not fall on the absorbing state. In this paper, we have focused on the stochastic simulation algorithms that provide statistically exact sequences of states, which is also fundamental to sample the quasistationary distribution. Therefore, our work contributes indirectly to this line of study.

Despite the fact that we have considered explicitly Markovian spreading processes, our composition and rejection scheme can be directly applied to certain classes of non-Markovian processes with completely monotone survival function, using the Laplace transform, in the spirit of Ref.~\cite{Masuda2018}. It can also directly be used with a variety of spreading processes on time-varying networks, where the structure evolves independently from the dynamics \cite{St-Onge2018,Taylor2012}, or co-evolves with it \cite{Gross2006,Scarpino2016}. Extension of this method for complex contagion \cite{Lehmann2018} would further be an interesting avenue.

Finally, from a more general perspective, we can argue that the idea behind composition and rejection is a systematic and efficient regrouping of independent processes, especially suited for multiscale propensities, as discussed in Sec.~\ref{sec:node-based}. It would be simple to exploit this scheme for many of the stochastic processes studied in network science, such as multistate dynamical processes \cite{Gleeson2013,Fennell2017} or network generation models \cite{Krapivsky2000,Hebert-Dufresne2011}.

\section*{Acknowledgments}
We thank Petter Holme for useful comments. We acknowledge Calcul Qu\'{e}bec for computing facilities. This research was undertaken thanks to the financial support from the Natural Sciences and Engineering Research Council of Canada (NSERC), the Fonds de recherche du Qu\'{e}bec --- Nature et technologies (FRQNT) and the Sentinel North program, financed by the Canada First Research Excellence Fund.

\appendix

\section{Some results for uncorrelated random networks \label{app:random_network}}

Over the last two decades, many analytical results have been obtained for spreading processes on uncorrelated random networks with an arbitrary degree distribution $P(k)$, i.e. the probability of having an edge with endpoint nodes of degree $k,k'$ is $P(k,k') \propto kk'P(k)P(k')$. To generate such uncorrelated networks, one must respect the structural cut-off \cite{Catanzaro2005}, i.e. $k_\mathrm{max} < \avg{k} N^{1/2}$. 

In the infinite size limit, threshold estimates $\lambda_c$ are obtained for spreading processes. For the SIS model, a good approximation is \cite{St-Onge2018}
\begin{align}\label{eq:threshold_SIS}
	\lambda_c^\mathrm{SIS} &= \mathrm{min} \pr{\sqrt{\frac{2}{k_\mathrm{max}}},\frac{\avg{k}}{\avg{k^2}-\avg{k}}} \;,
\end{align}
while for the SIR model, the exact result is \cite{Volz2008,Miller2011}
\begin{align}\label{eq:threshold_SIR}
	\lambda_c^\mathrm{SIR} &= \frac{\avg{k}}{\avg{k^2}-2\avg{k}} \;.
\end{align}
Equations \eqref{eq:threshold_SIS} and \eqref{eq:threshold_SIR} were used in the simulations of Fig.~\ref{fig:computation_time_comparison}.

Using the heterogeneous mean-field theory \cite{Boguna2002,Moreno2002}, we can also approximate the degree distribution of infected nodes near the phase transition. On the one hand, for the SIS model in the stationary state, we have
\begin{align}\label{eq:mean_field_SIS}
	P(k|I)^* \sim \frac{\lambda \theta^* k}{\mu + \lambda \theta^* k} P(k) \;,
\end{align}
with degree distribution $P(k)$, and where $\theta^*$ is the average fraction of infected neighbors for a susceptible node. Near the phase transition, $\theta^* \to 0$, which means that the average degree of infected nodes is
\begin{align}
	\avg{k|I}^* \sim \avg{k^2} \;.
\end{align}
On the other hand, if we consider a complete sequence of the SIR model, the average degree of recovered nodes is a good proxy, which is also $\sim\avg{k^2}$ near the phase transition. For power-law degree distribution $P(k) \sim k^{-\gamma}$, maximal degree $k_\mathrm{max} < \avg{k} N^{1/2}$ and $2 \leq \gamma < 3$, the second moment of the degree distribution scales with the number of nodes as
\begin{align}
	\avg{k^2} \sim \int_\mathrm{k_\mathrm{min}}^{k_\mathrm{max}} k^{2-\gamma} \dx k \sim N^{(3-\gamma)/2} \;.
\end{align}

\section{Overhead factor due to phantom events \label{sec:phantom_event_node-based}}

An element that we have discarded in our complexity analysis is the average fraction of \emph{phantom processes} \cite{Cota2017} $\phi$, i.e. processes $u_\alpha$ that do not change the state $\gv{x}$. One example is the transmission to an already infected node. It is assumed in our analysis of Sec.~\ref{sec:simulation_algorithms} that $\phi$ is upper-bounded by a constant $C < 1$, independent of the number of nodes in the system---this is well supported by the results of the composition and rejection scheme in Fig.~\ref{fig:computation_time_comparison}. It is worth pointing that the results of Fig.~\ref{fig:computation_time_comparison} are unbiased, since we counted the number of real transitions $\gv{x} \mapsto \gv{x}'$ with $\gv{x}' \neq \gv{x}$ to evaluate the empirical average computation time.

\begin{figure}[tb]
	\centering
	\includegraphics[width = 0.5\textwidth]{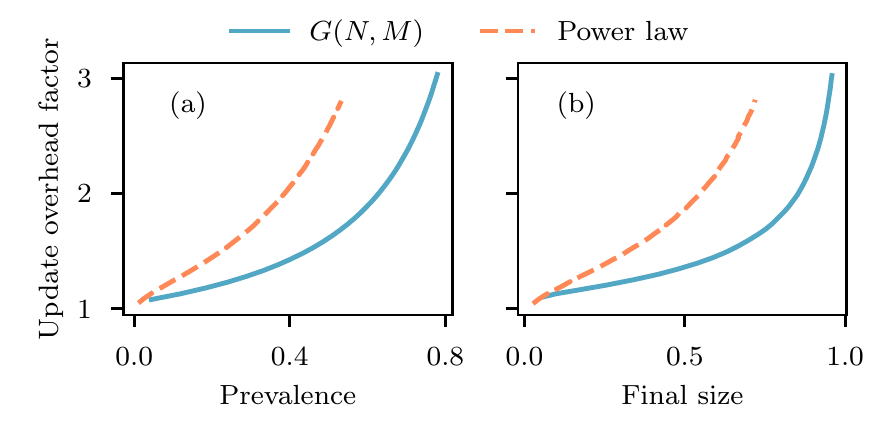}
	\caption{Expected overhead factor for the update time of a state due to phantom events, using node-based methods. This factor is estimated by $1/(1-\phi)$ and $\phi$ is measured for \textbf{(a)} the SIS model in the stationary state and \textbf{(b)} complete sequence of the SIR model. We sampled over 10 realizations of random graphs ensembles : the $G(N,M)$ ensemble with $N = 10^5$ and $M = 5\times 10^5$ ; random graphs with an expected degree sequence, a power law expected degree distribution $P(\kappa) \sim \kappa^{-\gamma}$ with $\gamma = 2.5$, a natural cut-off $k_\mathrm{max} < N$ and $N = 10^5$.}
	\label{fig:update_overhead_factor}
\end{figure}

For different values of prevalence (SIS) or final size (SIR), phantom processes can lead to a certain overhead. We show in Fig.~\ref{fig:update_overhead_factor} the expected multiplicative factor on the number of operations required for an update of the state, due to a certain fraction $\phi$ of phantom processes. 

Near the phase transition, the overhead factor is negligible, but it can get important for prevalence or final size near 1 where infected nodes are mostly surrounded by infected or recovered nodes. It would be possible to reduce the fraction of phantom processes $\phi$ in the node-based methods for the SIR model : we could count the number $\widetilde{k}_i \leq k_i$ of susceptible neighbors for a newly infected node, and modify the propensity to
\begin{align}
	\widetilde{\omega}_i = \delta(x_i,I)(\lambda \widetilde{k}_i + \mu) \;.
\end{align}
However, the update involving the infection of a degree $k_I$ node would now require
\begin{align}
	\mathcal{O}[\log(\log N) + k_I] \;,
\end{align}
which could be worst in certain cases according to our study. Since $\mathcal{O}[\log(\log N)]$---or $\mathcal{O}(1)$ for upper-bounded maximal degree---is a small number of operations, it is probably safer (and simpler) to keep the schemes introduced in Sec.~\ref{sec:node-based}.

\nocite{Tange2011a}

\section*{References}


%

\clearpage

\includepdf[pages=-]{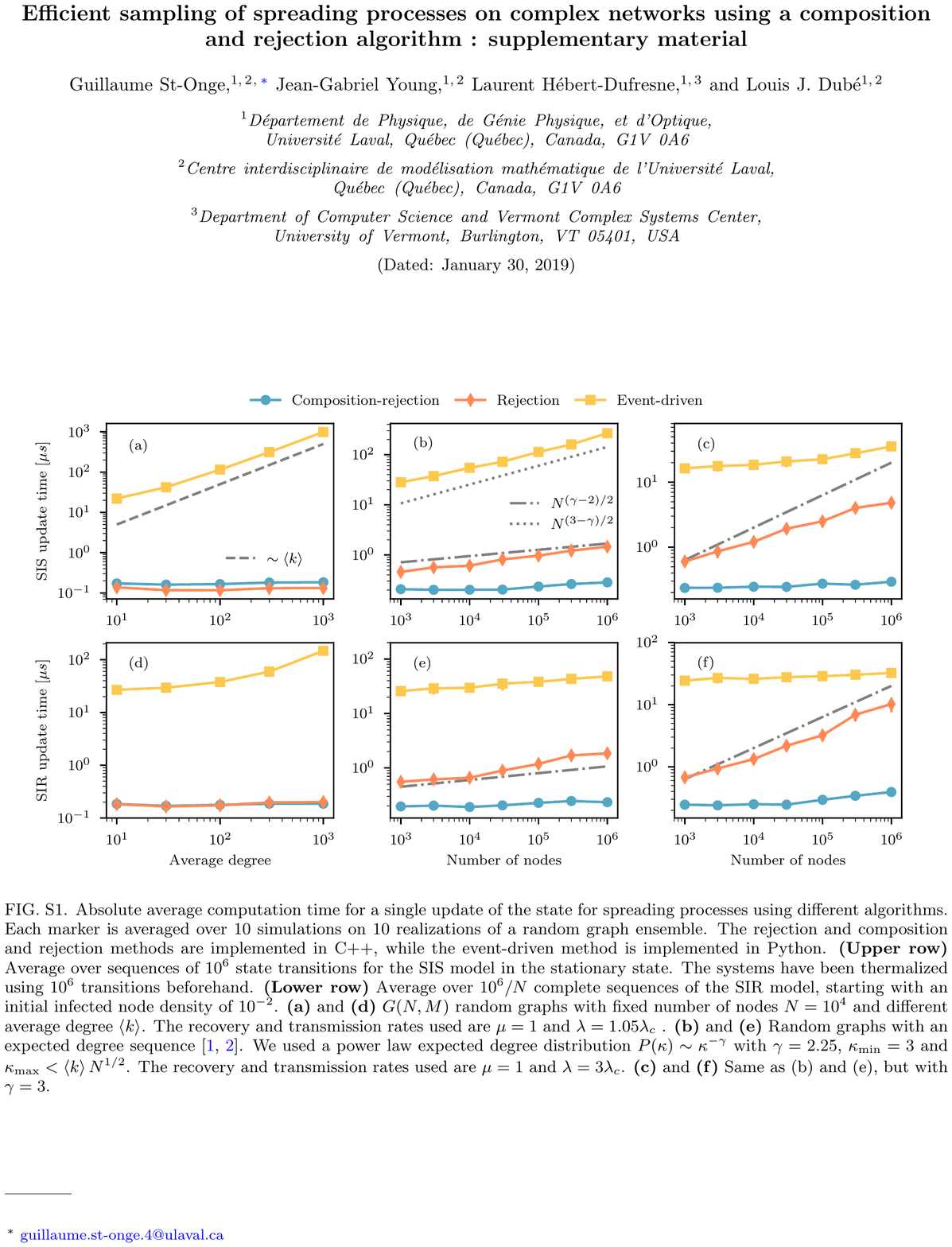}

\end{document}